\documentclass[a4paper,twoside]{article}
\newcommand{\affil}[1]{$^{\rm #1}$}
\usepackage[authoryear]{natbib}
\bibpunct{(}{)}{;}{a}{}{,}
\usepackage{graphicx}
\date{} %Please leave the date blank
\title{\large\bf\flushleft New High Precision Measurement of the
Reaction Rate of the ${}^{18}{\rm O}(p,\alpha){}^{15}{\rm N}$ Reaction via THM}
\author{\parbox{\textwidth}{\flushleft
\vspace{-0.5cm}
{\it M. La Cognata\affil{A,B}, C. Spitaleri\affil{A,B,C},
A.M. Mukhamedzhanov\affil{D}, B. Irgaziev\affil{E}, R.E. Tribble\affil{D},
A. Banu\affil{D}, S. Cherubini\affil{A,B}, A. Coc\affil{F}, V. Crucill\`a\affil{A,B},
V.Z. Goldberg\affil{D}, M. Gulino\affil{A,B}, G.G. Kiss\affil{G}, L. Lamia\affil{A,B}, Li Chengbo\affil{A,H},
J. Mrazek\affil{G}, R.G. Pizzone\affil{A,B}, S.M.R. Puglia\affil{A,B}, G.G. Rapisarda\affil{A,B},
S. Romano\affil{A,B}, M.L. Sergi\affil{A,B}, G. Tabacaru\affil{D}, L. Trache\affil{D},
W. Trzaska\affil{I} and A. Tumino\affil{A,B}}\\
\vspace{0.4cm}
{\small \affil{A}\,INFN - Laboratori Nazionali del Sud, Catania, Italy}\\
{\small \affil{B}\,DMFCI Universit\`a di Catania, Catania, Italy}\\
{\small \affil{D}\,Cyclotron Institute Texas A\&M University, College Station (TX), USA}\\
{\small \affil{E}\,GIK Institute of Engineering Sciences and Technology, Topi District, Swabi, Pakistan}\\
{\small \affil{F}\,CSNSM CNRS/IN2P3 Universit\`e Paris Sud, Orsay, France}\\
{\small \affil{F}\,ATOMKI, Debrecen, Hungary}\\
{\small \affil{G}\,Nuclear Physics Institute of ASCR, Rez near Prague, Czech Republic}\\
{\small \affil{H}\,China Institute of Atomic Energy, Beijing, China}\\
{\small \affil{I}\,Physics Department, University of Jyvaskyla, Finland}\\
{\small \affil{C}\,Email: Spitaleri@lns.infn.it}}}
\begin{document}
\twocolumn[
\begin{changemargin}{.8cm}{.5cm}
\begin{minipage}{.9\textwidth}
\vspace{-1cm}
\maketitle
%
%
%%%%%%%%%%%%%     ABSTRACT    %%%%%%%%%%%%%
%Abstract of no more than 200 words here.
\small{\bf Abstract:}
The $^{18}{\rm O}(p,\alpha)^{15}{\rm N}$ reaction rate has been extracted
by means of the Trojan horse method. For the first time the contribution
of the 20 keV peak has been directly evaluated, giving a value about 35\%
larger than previously estimated. The present approach has allowed
to improve the accuracy of a factor 8.5, as it is based on the measured
strength instead of educated guesses or spectroscopic measurements. The
contribution of the 90 keV resonance has been determined as well,
which turned out to be of negligible importance to astrophysics.

%%%%%%%%%%%%%     KEYWORDS    %%%%%%%%%%%%%
\medskip{\bf Keywords:} Write keywords here
% Please write all keywords in lower case. PASA uses the
% standard list of subject headings adopted by The Astrophysical Journal
% and available from http://www.journals.uchicago.edu/ApJ/keywords_text.html.
% Keywords are separated by em-dashes, i.e. ---

%%%%%%%%DO NOT EDIT%%%%%%%%%%%%
\medskip
\medskip
\end{minipage}
\end{changemargin}
]
\small
%%%%%%%%EDIT FROM HERE%%%%%%%%%%%%

\section{Astrophysical motivations}

Fluorine is one of the few elements whose nucleosynthesis is still
uncertain. Three possible astrophysical factories for fluorine
production have been identified, namely Type II Supernovae
(SNe II), Wolf-Rayet (WR) stars, and asymptotic giant branch (AGB)
stars \citep{REN04}. As regards AGB stars, which represents the final
nucleosynthetic phase in low and intermediate mass stars, spectroscopic observations
have shown that in giant stars of type K, M, MS, S, SC, and C fluorine
abundance is enhanced with respect to the solar by up to a factor 30
\citep{JOR92}. Thus low-mass evolved stars are observationally confirmed
astrophysical sites where fluorine is produced. Inside AGB stars, $^{19}$F nucleosynthesis
takes place at the same evolutionary stage and in the same region
as the s-process nucleosynthesis, which represents the nuclear
process leading to the generation of heavy elements along the
stability valley. For these reason AGB stars play an extremely
important role in astrophysics and the understanding of fluorine
production, allowing to constrain the existing models \citep{LUG04},
would make predictions on AGB star nucleosynthesis and s-process
element yields more accurate.
In detail, $^{19}$F is produced during the thermal
pulse that is ignited in the $^4$He-rich intershell region of AGB stars,
following the ingestion of the $^{13}$C pocket. The subsequent
third dredge-up (TDU) episode mixes the products of shell flash
He-burning (thermal pulse), including fluorine, and $s$-process nuclei to the outer layers.
Because $^{19}$F abundance is very sensitive to the temperatures and the mixing
processes taking place inside AGB stars, it constitutes a key parameter to constrain
AGB star models \citep{LUG04}. Anyway, if standard theoretical
abundances are compared to the observed ones \citep{JOR92}, a remarkable discrepancy
shows up because the largest $^{19}$F abundances cannot be matched
for the typical $^{12}{\rm C}/^{16}{\rm O}$ ratios \citep{LUG04}.
It has been shown that extra-mixing phenomena, such as the cool bottom
process \citep{NOL03}, could help to pin down the origin of this
discrepancy \citep{LUG04}.

A complementary way to explain $^{19}$F abundance
can be provided by nuclear physics, in particular by an improved
measurement of the $^{18}$O$(p,\alpha){}^{15}$N reaction rate.
In fact this reaction represents the main $^{15}$N production channel, which
is burnt to $^{19}$F via the $^{15}$N($\alpha$,$\gamma$)$^{19}$F reaction
during thermal pulses, at temperatures of the order of 10$^8$ K.
Thus a larger $^{18}$O$(p,\alpha){}^{15}$N reaction rate
would lead to an increase of the $^{19}$F supply, while the $^{12}{\rm C}/^{16}{\rm O}$
ratio would not change. Such an alternative account would also imply
an enrichment of $^{15}$N in the stellar surface, as a result of
the cool bottom processing of material from AGB outer layers at the bottom
of the convective envelope \citep{NOL03}, at temperatures of about 10$^7$ K.
Therefore a new investigation of the $^{18}$O$(p,\alpha){}^{15}$N reaction at
low energies, in the 0-1 MeV energy range would also play a key role
to explain the long-standing problem of the $^{14}{\rm N}/^{15}{\rm N}$
ratio in meteorite grains (\cite{NOL03} and references therein) besides
the $^{19}$F yield. Indeed this ratio turns out to be much smaller than
the predicted one for mainstream and A+B grains and any
proposed astrophysical explanation, including extra-mixing scenarios,
could not help the make the model predictions more accurate \citep{NOL03}.
In the following the first measurement of the low-laying resonances
in the $^{18}$O$(p,\alpha){}^{15}$N reaction is discussed and
how the reaction rate is influenced is extensively illustrated.

\section{Current status}

In the 0-1 MeV energy range, which is the most relevant to astrophysics,
9 resonances show up in the $^{18}$O$(p,\alpha){}^{15}$N cross section.
Among these, the 20 keV, 144 keV and the 656 keV resonances determine the reaction rate \citep{ANG99}.
Though these resonances has been the subject of
several direct experimental investigations \citep{MAK78,LOR79} as well as of many
spectroscopic studies \citep{YAG62,CHA86,WIE80,SCH70}, the reaction rate for
this process has a considerable uncertainty \citep{ANG99}.
With regard to the 20 keV resonance, its strength is known only
from spectroscopic measurements performed through the transfer reaction
$^{18}{\rm O}(^{3}{\rm He},d)^{19}{\rm F}$ \citep{CHA86}
and the direct capture reaction $^{18}{\rm O}(p,\gamma)^{19}{\rm F}$
\citep{WIE80}. Therefore the deduced reaction rate is affected by large and
not-well-defined uncertainties, because the deduced strengths
are strongly model dependent. In fact they rely on the optical model potentials
adopted in the data analysis, and different set of potentials or of parameters,
though giving a reasonable account of the experimental data, lead
to the extraction of different spectroscopic factors.
An additional important source of uncertainty on the reaction rate
is connected with the determination of the resonance energy for this resonance \citep{CHA86}.
The resonance at 143.5 keV is fairly well established \citep{LOR79}. The broad resonance at
656 keV gives strong contribution both at low and high temperatures. The total width of
this high energy resonance is badly known and, as a consequence, also its
contribution to the reaction rate. Two sets of widths are present in the literature, namely
\cite{YAG62} and \cite{LOR79}. To sum up, the uncertainties on nuclear physics inputs have made
astrophysical predictions far from conclusive \citep{NOL03}. As already discussed,
in this paper we will focus on the low-laying resonances, below about 200 keV.
In this range an additional resonance at 90 keV in the
$^{18}$O$(p,\alpha){}^{15}$N cross section, corresponding to the
8.084 MeV excited state in $^{19}$F occurs. The influence of this
level on the reaction rate is also established.

\section{The Trojan horse method}

Two main reasons make the direct measurement of the cross section
of astrophysically relevant reaction not accurate or even impossible:
on one hand the presence of the Coulomb barrier, exponentially suppressing
the cross section at the lowest energies, on the other the presence
of atomic electrons. As regards the Coulomb suppression, inside the Gamow
energy window the cross section for reactions among charged particle
drops well below 10$^{-12}$ barn, thus making statistical accuracy
and signal-to-noise ratio very poor. Even in the few cases where
the measurement has been possible, especially in the case of light nuclei,
thanks to improved techniques and underground laboratories \citep{FIO95},
the presence of atomic electrons has prevented the access to the relevant
information, that is the bare nucleus cross section. In fact atomic electrons
screen the nuclear charges thus determining an enhancement of the cross
section at the lowest energies, which is not related to nuclear
physics \citep{ASS87}. Therefore the cross section at the energies relevant to astrophysics
has to be extracted by means of extrapolation from higher energies,
where the cross section is more easily measured. The extrapolation is
worked out by means of R-matrix calculations (see, for instance,
\cite{BAR02}) or, if no calculations are available, by means of
simple polynomial fit. As a result large uncertainties can be introduced
into the astrophysical models because of an incorrect estimate of the relevant
reaction rates, as we have argued in the case of the $^{18}$O$(p,\alpha){}^{15}$N reaction.

\begin{figure}[t]
\begin{center}
\includegraphics[scale=0.4, angle=0]{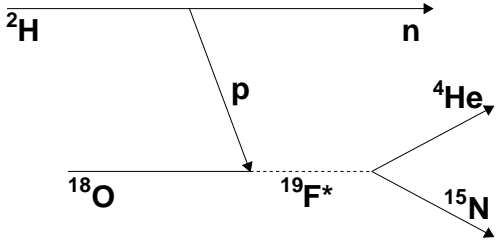}
\caption{Simple sketch of the $^2{\rm H}(^{18}{\rm O},\alpha{}^{15}{\rm N})n$ TH reaction.}
\label{thmgraph}
\end{center}
\end{figure}

In order to reduce the nuclear uncertainties affecting its
reaction rate we have performed an experimental study of
the $^{18}$O$(p,\alpha){}^{15}$N reaction by means of the Trojan
horse method (THM), which is an indirect
technique to measure the relative energy-dependence of a charged-particle reaction cross
section at energies well below the Coulomb barrier
(\cite{LAC07,SPI99} and references therein). The cross section of
the $^{18}$O$(p,\alpha){}^{15}$N reaction is deduced from the
$^2$H($^{18}$O$,\alpha{}^{15}$N)$n$ three-body process, performed in quasi-free (QF)
kinematics. The beam energy is chosen larger than the Coulomb
barrier for the interacting nuclei, so the break-up of the deuteron
(acting as the $Trojan$-$horse$ nucleus) takes place inside the $^{18}$O
nuclear field. Therefore, the cross section of the $^{18}$O$(p,\alpha){}^{15}$N
reaction is not suppressed by the Coulomb interaction of the
target-projectile system, while no electron screening enhancement is
spoiling the nuclear information because the reaction is performed at
high energies (several tens of MeV). The QF reaction mechanism for the
$^2$H($^{18}$O$,\alpha{}^{15}$N)$n$ process is sketched in Fig. \ref{thmgraph}.

The THM cross section for the ${}^{18}{\rm O}+d(p \oplus n) \to {}^{15}{\rm N}+\alpha+n$ reaction
proceeding through a resonance ${}^{19}{\rm F}_{i}$ in the subsystem ${}^{19}{\rm F} = {}^{18}{\rm O} + p
= {}^{15}{\rm N} + \alpha$ can be obtained if the process is described as a transfer to the
continuum, where the emitted neutron keeps the same momentum as the one it has inside deuteron
(QF condition). If such a hypothesis is satisfied, the cross section for the QF
$^2$H($^{18}$O$,\alpha{}^{15}$N)$n$ three-body reaction is \citep{LAC07,muk2007}
\begin{equation}
\frac{d^2\sigma}{dE_{\alpha{}^{15}{\rm N}}\,d\Omega_n} \propto
\frac{\Gamma_{(\alpha{}^{15}{\rm N})_{i}}(E)\,|M_{i}(E)|^{2}
}{(E - E_{R_{i}})^{2} + \Gamma_{i}^{2}(E)/4} .
\label{d3sigma}
\end{equation}
Here, $M_{i}(E)$ is the direct transfer reaction amplitude for the
binary reaction ${}^{18}{\rm O}+d \to {}^{19}{\rm F}_i+n$
leading to the population of the i-th resonant state of
$^{19}$F with the resonance energy $E_{R_{i}}$,  $E$ is the ${}^{18}{\rm O} - p$
relative kinetic energy related to $E_{{}^{15}{\rm N} - \alpha}$ by
the energy conservation law, $\Gamma_{(\alpha{}^{15}{\rm N})_{i}}(E)$ is the partial
resonance width for the decay $^{19}{\rm F}_{i} \to \alpha - {}^{15}{\rm N}$ and $\Gamma_{i}$ is the total
resonance width of $^{19}{\rm F}_{i}$. The appearance of the transfer reaction amplitude
$M_{i}(E)$ instead of the entry channel partial resonance width $\Gamma_{(p{}^{18}{O})_{i}}(E)$
is the main difference between the THM cross section and the cross section for the
resonant binary sub-reaction ${}^{18}{\rm O} + p \to {}^{15}{\rm N} + \alpha$ \citep{LAC07,muk2007}.
Therefore the cross section of the three-body process can be easily connected to
the one for the two-body reaction of interest by evaluating the transfer
amplitude $M_{i}(E)$. In the plane wave approximation
$M_{i} \approx  \varphi_{d}(p_{pn})\,W_{p{}^{18}{\rm O}}({\rm {\bf p}}_{p{}^{18}{\rm O}})$,
where $\varphi_{d}(p_{pn})$ is the Fourier transform
of the $s$-wave radial $p-n$ bound-state wave function, $p_{pn}$ is the $p-n$ relative momentum, and
$W_{p{}^{18}{\rm O}}({\rm {\bf p}}_{p{}^{18}{\rm O}})$ is the form factor for the synthesis
${}^{18}{\rm O} + p \to {}^{19}{\rm F}_{i}$ \citep{LAC07,LAC08}.
In the present case no distortion are observed because of the
high beam energy and because the emitted neutron in the exit
channel has no long range Coulomb interaction.

\section{Experimental investigation}

The experiment was performed at Laboratori Nazionali del Sud, Catania (Italy)
and represents the continuation of the one carried out at the Cyclotron
Institute, Texas A\&M University, Texas (USA).
The SMP Tandem Van de Graaff accelerator provided the 54 MeV $^{18}$O beam
which was accurately collimated to achieve the best angular resolution.
The intensity was 5 enA on the average and the
relative beam energy spread was about 10$^{-4}$. Thin
self-supported deuterated polyethylene (CD2) targets, about
100 $\mu g/cm^{2}$ thick, were adopted in order to minimize
angular straggling. The detection setup consisted of a telescope (A),
to single out Z=7 particles, made up of an ionization chamber and
a silicon position sensitive detector (PSD A). Negligible angular straggling
was introduced on the $^{15}$N detection by the ionization chamber.
Three additional silicon PSD's (B, C and D) were placed on the opposite side,
with the aim of detecting alpha particles from the $^2$H($^{18}$O$,\alpha{}^{15}$N)$n$
QF three-body process. No $\Delta E$ detectors were put in front of PSD's B, C
and D to decrease detection thresholds and to achieve the
best energy and angular resolution. Angular conditions were selected
in order to maximize the expected QF contribution.

A description of the data analysis is reported in \cite{LAC08},
here we shortly summarize the main stages.
After detector calibration, the first step of the analysis was the
reaction channel selection. This is necessary because several
reactions can take place in the target, while only partial
particle identification is allowed by the experimental setup. In detail,
$\alpha$-particle identification as well as A=15 selection in PSD A
were accomplished from the kinematics of the events.
Indeed in a three-body reactions the events gather in
some well-defined kinematical regions, fixed by the
Q-value of the three-body process. The procedure discussed
in \cite{COS90} was then applied after gating on the time-to-amplitude
converter to select the coincidence peak and on the $\Delta E - E$ 2D
spectra to select the nitrogen locus. The kinematic locus of
the $^2$H($^{18}$O,$\alpha$$^{15}$N)$n$ reaction was then extracted
and compared to the corresponding one, obtained by
means of a Monte Carlo simulation, showing that no additional channels
contribute to the experimental kinematic locus.

A further study on reaction dynamics is necessary to select
those kinematic regions where QF break-up is dominant and can be separated
from direct break-up (DBU) or sequential decay (SD).
To this purpose the $E_{^{15}{\rm N}-n}$ and the $E_{\alpha-n}$
relative energy spectra were extracted to
evaluate the contribution from $^{16}$N$^*$ and $^5$He$^*$
excited states. On the other hand, since the same resonances
in the the $^{19}{\rm F}^*+n$ channel can be
observed through QF and SD reaction mechanisms, the
experimental neutron momentum distribution has been evaluated. Indeed,
only if the deuteron break-up process is direct the
neutron momentum distribution keeps the same shape as inside $d$.
The procedure to extract the experimental neutron momentum
distribution is extensively discussed in \cite{LAC07} and \cite{SPI04}.
The resulting distribution is compared with the theoretical one
given by the square of the Hulth\'en wave function
in momentum space \citep{LAC07,SPI04}. The
good agreement demonstrates that the QF mechanism is present and
dominant in the $p_3<50$ MeV/c neutron momentum range. In addition, inside
this region the contribution from the SD of $^{16}$N excited
states is negligible. For these reasons, in the following analysis
only the phase space region for which the $p_3<50$ MeV/c condition
is satisfied is taken into account.

\begin{figure}[t]
\begin{center}
\includegraphics[scale=0.4, angle=0]{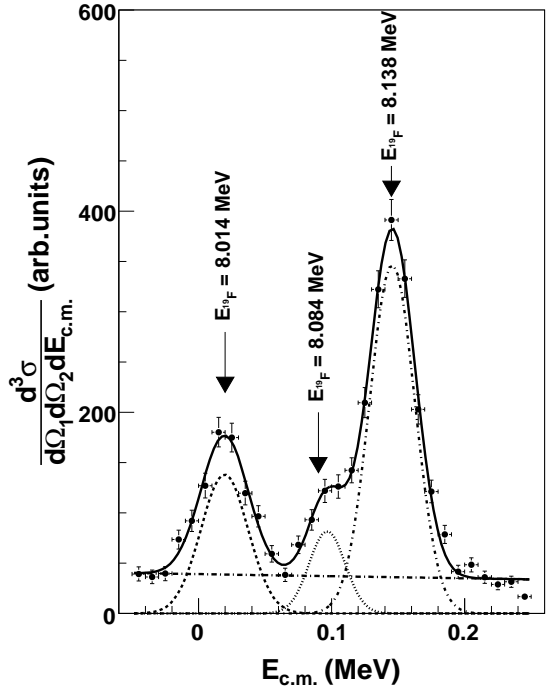}
\caption{Cross section of the $^2{\rm H}(^{18}{\rm O},\alpha{}^{15}{\rm N})n$ TH reaction.
See text fo details.}
\label{cs18}
\end{center}
\end{figure}

\begin{figure}[t]
\begin{center}
\includegraphics[scale=0.4, angle=0]{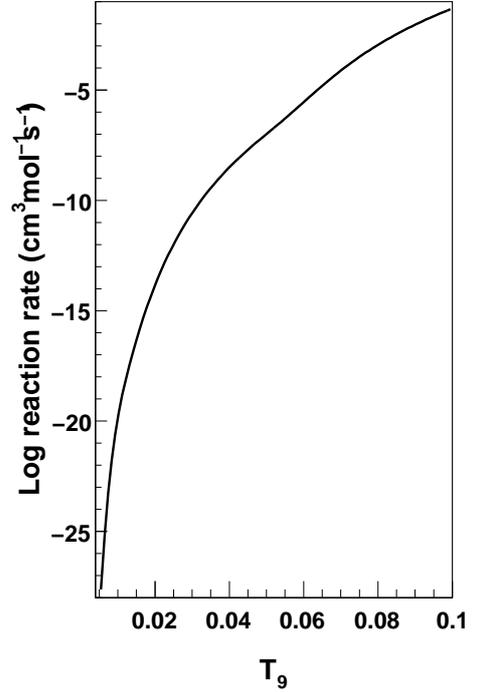}
\caption{Reaction rate of the ${}^{18}{\rm O}(p,\alpha){}^{15}{\rm N}$ reaction.}
\label{rate18}
\end{center}
\end{figure}

The extracted three-body cross section
has been integrated in the whole angular range. The
resulting $^{2}{\rm H}(^{18}{\rm O},\alpha^{15}{\rm N})n$ reaction
cross section is shown in Fig. \ref{cs18} (full circles). The experimental
energy resolution turned out to be about 40 keV (FWHM). Horizontal
error bars represent the integration bin while the vertical ones arise
from statistical uncertainty and angular distribution
integration. The solid line in the figure is the sum
of three Gaussian functions to fit the resonant behavior and a straight line
to account for the non-resonant contribution to the cross section.
The resonance energies were then deduced: ${\rm E}_{R_1} = 19.5 \pm 1.1$ keV,
${\rm E}_{R_2} = 96.6 \pm 2.2$ keV and ${\rm E}_{R_3} = 145.5 \pm 0.6$ keV (in fair
agreement with the ones reported in the literature \citep{ANG99}) as
well as the peak values of each resonance in arbitrary units: $N_1 = 138 \pm 8$, $N_2 = 82 \pm 9$
and $N_3 = 347 \pm 8$. The peak values were used to derive
the resonance strengths:
\begin{equation}
(\omega\gamma)_i = \frac{2J_{{}^{19}{\rm F}_i}+1}{(2J_{{}^{18}{\rm O}}+1)(2J_p+1)}
\frac{\Gamma_{(p{}^{18}{\rm O})_{i}}\Gamma_{(\alpha{}^{15}{\rm N})_{i}}}{\Gamma_i} \, ,
\label{omegam}
\end{equation}
that are the relevant parameters for astrophysical application
in the case of narrow resonances \citep{ANG99}.
The peak THM cross section taken at the $E_{R_{i}}$  resonance
energy for the $(p,\alpha)$ reaction $A + x \to C + c$ is given by
\begin{equation}
N_i =  4\,\frac{\Gamma_{\alpha_{i}}(E_{R_i})\,M_{i}^2(E_{R_i})}{\Gamma_{i}^{2}\,
(E_{R_{i}})},
\label{maxval}
\end{equation}
where $\Gamma_{(\alpha{}^{15}{\rm N})_{i}}(E) \equiv \Gamma_{\alpha_{i}}(E)$.
In this work we did not extract the absolute value of the cross section. Anyway the proton and
alpha partial widths for the third resonance are well known \citep{ANG99}, thus we can determine
the strength for the 20 keV and 90 keV resonances from the ratio of the peak
values of the THM cross sections, as discussed by \cite{LAC08}.
The electron screening gives a negligible contribution around 144 keV
(4\% maximum \citep{ASS87}), thus no systematic uncertainty is introduced by
normalizing to the highest energy resonance. If $(\omega\gamma)_3$ is taken from \cite{BEC95},
one gets $(\omega\gamma)_1=8.3^{+3.8}_{-2.6}\times 10^{-19}$ eV,
which is well within the confidence range established by NACRE, $6^{+17}_{-5}\times 10^{-19}$ eV \citep{ANG99}.
This is because NACRE recommended value is based on spectroscopic data while the present result
is obtained from experimental ones, thus increasing the accuracy of the deduced resonance
strength. The largest contribution to the error is due to the uncertainty on the resonance energy,
while statistical and normalization errors sum up to about 9.5\%.
To cross check the method, we have extracted the resonance strength of the 90 keV resonance,
which is known with fairly good accuracy ($(1.6 \pm 0.5)\times 10^{-7}$ eV \citep{ANG99}).
We got $(\omega\gamma)_2=(1.76 \pm 0.33)\times 10^{-7}$ eV (statistical and
normalization errors $\sim$ 13\%), in good agreement with the strength given by NACRE,
giving us confidence in the theory used in the present paper.

\section{Extraction of the reaction rate}

By using the narrow resonance approximation \citep{ANG99}, which is fulfilled for the resonances under investigation,
the reaction rate for the ${}^{18}{\rm O}(p,\alpha){}^{15}{\rm N}$ reaction has been deduced.
According to this approximation, the contribution to the rate of the i-th resonance is given by:
\begin{eqnarray}
N_A\left< \sigma v \right>_{R_i} = \nonumber\\
N_A \left( \frac{2\pi}{\mu k_B} \right)^{3/2} \hbar^2
( \omega\gamma )_i T^{-3/2} \exp\left( -E_{R_i}/k_B T \right)
\label{ratefori}
\end{eqnarray}
where $\mu$ is the reduced mass for the projectile-target system and $T$ is the temperature
of the astrophysical site. The resulting rate $R_{{}^{18}{\rm O}(p,\alpha){}^{15}{\rm N}}$
is displayed, as a function of the temperature,
in Fig. \ref{rate18}. The analytic expression of the reaction rate (with $\approx$ 10\% accuracy) is:
\begin{eqnarray}
R_{{}^{18}{\rm O}(p,\alpha){}^{15}{\rm N}} = \nonumber \\
\frac{5.58\,10^{11}}{T_9^{2/3}}\,\exp\left(-\frac{16.732}{T_9^{1/3}}-\left(\frac{T_9}{0.51}\right)^2\right)\,\nonumber \\
(1 + 3.2\,T_9 + 21.8\,T_9^2) + \frac{1.375\,10^{-13}}{T_9^{3/2}}\,\exp\left(-\frac{0.232}{T_9}\right)\nonumber \\
+ \frac{2.58\,10^4}{T_9^{3/2}}\,\exp\left(-\frac{1.665}{T_9}\right)+ \frac{3.24\,10^8}{T_9^{0.378}}\,\exp\left(-\frac{6.395}{T_9}\right)
\label{anarate}
\end{eqnarray}
where $T_9$ is the temperature in billion kelvin and the reaction rate $R_{{}^{18}{\rm O}(p,\alpha){}^{15}{\rm N}}$
is measured in cm$^3$mol$^{-1}$sec$^{-1}$. This expression is obtained by using as a fitting function
a formula similar to the NACRE one, leaving as free parameters the numerical coefficients and using
as initialization parameters the NACRE ones \citep{ANG99}.

\begin{figure}[t]
\begin{center}
\includegraphics[scale=0.36, angle=0]{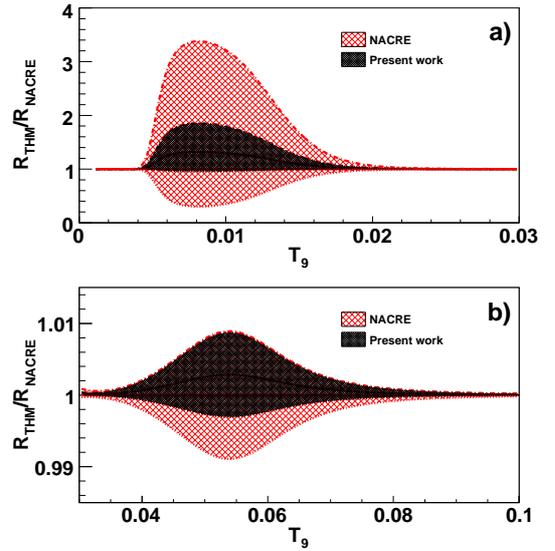}
\caption{Comparison of the reaction rate of the ${}^{18}{\rm O}(p,\alpha){}^{15}{\rm N}$
reaction with the NACRE one \citep{ANG99}.}
\label{cfrrate}
\end{center}
\end{figure}

Because of the strong dependence on the temperature (for a factor 10 change in
the temperature, the reaction rate increases by about 30 orders of magnitude), any comparison of
the present results with the one in the literature is very difficult. In order
to compare with the one reported in NACRE
\citep{ANG99}, the ratio of the THM reaction rate to
the NACRE one for the $^{18}{\rm O}(p,\alpha){}^{15}{\rm N}$ reaction is shown as a full black
line in Fig. \ref{cfrrate}. In this representation, the NACRE rate is given by a full
red line, that is by 1 in the whole examined range.
The dot-dashed and dotted black lines represent the upper and lower limits respectively, allowed
by the experimental uncertainties. As before, black and red lines mark THM and NACRE data.
In the low temperature region (below
T$_9=0.03$, Fig. \ref{cfrrate}a) the reaction rate can be about 35\% larger
than the one given by NACRE, while the indetermination is greatly reduced with respect
to the NACRE one, by a factor $\approx$ 8.5, in the case the error on the NACRE rate
is supposed to come entirely from the uncertainty on the 20 keV resonance strength,
to make the comparison homogenous.
Those temperatures are typical of the bottom of the convective envelope, thus an increase of
this reaction rate might have important consequences on the cool bottom process
\citep{NOL03} and, in turn, on the surface abundances and isotopic ratios in AGB stars.
The 8.084 MeV excited state of $^{19}$F (corresponding to the 90 keV resonance) provides a negligible
contribution to the reaction rate in agreement with the
previous estimate by \cite{CHA86}. This is clearly displayed by Fig. \ref{cfrrate}b),
where an increase of less than 1\% is obtained due to the THM measurement
of the 90 keV level resonance strength.
For completeness, the THM reaction rate and the NACRE one are given in Tab. \ref{tablerate},
together with the upper and lower limits allowed by experimental uncertainties. As discussed
before, the confidence range for the NACRE rate is evaluated by assuming that the only
source of indetermination is coming from the 20 keV resonance, to make the
comparison meaningful.

\begin{table*}[t]
\begin{center}
\caption{Rate of the ${}^{18}{\rm O}(p,\alpha){}^{15}{\rm N}$ reaction,
in comparison with the one from NACRE \citep{ANG99}}\label{tablerate}
\begin{tabular}{lcccccc}
\hline
Temperature ($10^9$ K) & \multicolumn{3}{c}{Rate THM (cm$^3$mol$^{-1}$s$^{-1}$)} & \multicolumn{3}{c}{Rate NACRE (cm$^3$mol$^{-1}$s$^{-1}$)} \\
                       & lower & adopted & upper & lower & adopted & upper \\
\hline
0.007 & $8.12\,10^{-25}$ & $1.11\,10^{-24}$ & $1.54\,10^{-24}$ & $2.75\,10^{-25}$ & $8.44\,10^{-25}$ & $2.78\,10^{-24}$ \\
0.008 & $4.02\,10^{-23}$ & $5.55\,10^{-23}$ & $7.79\,10^{-23}$ & $1.25\,10^{-23}$ & $4.19\,10^{-23}$ & $1.42\,10^{-22}$ \\
0.009 & $8.60\,10^{-22}$ & $1.18\,10^{-21}$ & $1.65\,10^{-21}$ & $2.78\,10^{-22}$ & $8.95\,10^{-22}$ & $2.99\,10^{-21}$ \\
0.010 & $1.03\,10^{-20}$ & $1.39\,10^{-20}$ & $1.92\,10^{-20}$ & $3.71\,10^{-21}$ & $1.06\,10^{-20}$ & $3.42\,10^{-20}$ \\
0.011 & $8.15\,10^{-20}$ & $1.07\,10^{-19}$ & $1.45\,10^{-19}$ & $3.47\,10^{-20}$ & $8.43\,10^{-20}$ & $2.53\,10^{-19}$ \\
0.012 & $4.90\,10^{-19}$ & $6.22\,10^{-19}$ & $8.14\,10^{-19}$ & $2.52\,10^{-19}$ & $5.04\,10^{-19}$ & $1.36\,10^{-18}$ \\
0.013 & $2.45\,10^{-18}$ & $2.96\,10^{-18}$ & $3.72\,10^{-18}$ & $1.51\,10^{-18}$ & $2.50\,10^{-18}$ & $5.87\,10^{-18}$ \\
0.014 & $1.07\,10^{-17}$ & $1.24\,10^{-17}$ & $1.48\,10^{-17}$ & $7.76\,10^{-18}$ & $1.09\,10^{-17}$ & $2.17\,10^{-17}$ \\
0.015 & $4.30\,10^{-17}$ & $4.75\,10^{-17}$ & $5.40\,10^{-17}$ & $3.48\,10^{-17}$ & $4.35\,10^{-17}$ & $7.28\,10^{-17}$ \\
0.016 & $1.58\,10^{-16}$ & $1.69\,10^{-16}$ & $1.85\,10^{-16}$ & $1.39\,10^{-16}$ & $1.60\,10^{-16}$ & $2.30\,10^{-16}$ \\
0.018 & $1.72\,10^{-15}$ & $1.76\,10^{-15}$ & $1.83\,10^{-15}$ & $1.64\,10^{-15}$ & $1.72\,10^{-15}$ & $2.02\,10^{-15}$ \\
0.020 & $1.41\,10^{-14}$ & $1.42\,10^{-14}$ & $1.44\,10^{-14}$ & $1.38\,10^{-14}$ & $1.41\,10^{-14}$ & $1.50\,10^{-14}$ \\
0.025 & $1.00\,10^{-12}$ & $1.01\,10^{-12}$ & $1.01\,10^{-12}$ & $1.00\,10^{-12}$ & $1.00\,10^{-12}$ & $1.01\,10^{-12}$ \\
0.030 & $2.64\,10^{-11}$ & $2.64\,10^{-11}$ & $2.64\,10^{-11}$ & $2.64\,10^{-11}$ & $2.64\,10^{-11}$ & $2.64\,10^{-11}$ \\
0.040 & $3.12\,10^{-9}$  & $3.12\,10^{-9}$  & $3.12\,10^{-9}$  & $3.12\,10^{-9}$  & $3.12\,10^{-9}$  & $3.12\,10^{-9}$ \\
0.050 & $1.01\,10^{-7}$  & $1.01\,10^{-7}$  & $1.01\,10^{-7}$  & $1.01\,10^{-7}$  & $1.01\,10^{-7}$  & $1.01\,10^{-7}$ \\
0.060 & $2.81\,10^{-6}$  & $2.81\,10^{-6}$  & $2.81\,10^{-6}$  & $2.81\,10^{-6}$  & $2.81\,10^{-6}$  & $2.81\,10^{-6}$ \\
0.070 & $7.52\,10^{-5}$  & $7.52\,10^{-5}$  & $7.52\,10^{-5}$  & $7.52\,10^{-5}$  & $7.52\,10^{-5}$  & $7.52\,10^{-5}$ \\
0.080 & $1.10\,10^{-3}$  & $1.10\,10^{-3}$  & $1.10\,10^{-3}$  & $1.10\,10^{-3}$  & $1.10\,10^{-3}$  & $1.10\,10^{-3}$ \\
0.090 & $9.07\,10^{-3}$  & $9.07\,10^{-3}$  & $9.07\,10^{-3}$  & $9.07\,10^{-3}$  & $9.07\,10^{-3}$  & $9.07\,10^{-3}$ \\
0.100 & $4.88\,10^{-2}$  & $4.88\,10^{-2}$  & $4.88\,10^{-2}$  & $4.88\,10^{-2}$  & $4.88\,10^{-2}$  & $4.88\,10^{-2}$ \\
\hline
\end{tabular}
\medskip\\
%$^a$Table footnotes go here.\\
\end{center}
\end{table*}

\section{Final remarks}

In this paper we have evaluated the influence of the new improved
measurement of the 20 keV resonance on the reaction rate of the $^{18}{\rm O}(p,\alpha){}^{15}{\rm N}$
reaction. In fact, for the first time, the strength of the low-lying 20 keV resonance
in ${}^{19}{\rm F}$ has been experimentally determined thanks to the
use of the indirect THM, while the same measurements have been proved
elusive for any direct approach (see \cite{LAC08} for a detailed discussion).
The present result turns out to be about 35\% larger than
the NACRE rate \citep{ANG99} in the region where the effect
of the presence of the 20 keV resonance is more intense.
This newly developed approach, which is based on experimental
data in contrast to the NACRE one that relies on various
kinds of estimates, has allowed to enhance the accuracy of the rate,
reducing the uncertainty due to the poor knowledge of the
parameters of the 20 keV resonance in the $^{18}{\rm O}(p,\alpha){}^{15}{\rm N}$ reaction
by a factor $\approx 8.5$. Such a remarkable improvement is mainly
due to two reasons. On one hand, the THM bring to the determination of the strength
of the unknown resonance avoiding information about the spectroscopic factors,
which are a primary source of systematic errors. On the other, our results
are not affected by the electron screening, which can enhance the cross section
by a factor larger than about 2.4 at 20 keV \citep{ASS87}, thus spoiling any direct
measurement of this resonance. As a next step, the astrophysical
consequences of the present work are to be evaluated, both onto
the scenarios sketched in the introduction and on alternative
environments. In addition, at higher temperatures, higher energy resonances
in the $^{18}{\rm O}(p,\alpha){}^{15}{\rm N}$ reaction can play a role.
These studies will be the subject of forthcoming works.

%\end{multicols}

\end{document}